\documentclass[aps,reprint,prl,showpacs,superscriptaddress]{revtex4-1}

\usepackage{bm}
\usepackage{amssymb}
\usepackage{amsmath}
\usepackage{color}

\usepackage{graphicx,epstopdf}

\begin{document}
\title{Exploring the complex world of two-dimensional ordering with three modes}
\author{S. K. Mkhonta}
\affiliation{Department of Physics and Astronomy, Wayne State University,
Detroit, Michigan 48201, USA}
\affiliation{Department of Physics, University of Swaziland, Private Bag 4,
Kwaluseni, Swaziland}
\author{K. R. Elder}
\affiliation{Department of Physics, Oakland University, Rochester, Michigan
48309, USA}
\author{Zhi-Feng Huang}
\affiliation{Department of Physics and Astronomy, Wayne State University,
Detroit, Michigan 48201, USA}

\date{\today}

\begin{abstract}
The world of two-dimensional crystals is of great significance for the
design and study of structural and functional materials with novel properties.
Here we examine the mechanisms governing the formation and dynamics of these
crystalline or polycrystalline states and their elastic and plastic properties
by constructing a generic multi-mode phase field crystal model. Our results
demonstrate that a system with three competing length scales can order into
all five Bravais lattices, and other more complex structures including honeycomb,
kagome and other hybrid phases. In addition, non-equilibrium phase transitions
are examined to illustrate the complex phase behavior described
by the model. This model provides a systematic path to predict the influence of
lattice symmetry on both structure and dynamics of crystalline and defected
systems.
\end{abstract}


\maketitle

Two-dimensional (2D) crystalline materials have been of tremendous interest in
both fundamental research and technological applications due to their
extraordinary properties and functionalities that are absent in three-dimensional
materials. A typical and well-known example is graphene, which exhibits
exceptional electronic, mechanical and thermal properties \cite{Geim07,Pantelides12}.
Recent efforts have been extended to the search for and study of 2D monolayer sheets
of graphene type or beyond, such as group IV elements of silicene \cite{Vogt12}
and germanane (GeH) \cite{Bianco13}, BN and BNC \cite{Ci10}, and semiconducting
MoS$_2$ and MoSe$_2$ \cite{Tongay12}. On larger length scales much progress
has been made in the self-assembly of 2D crystals using particles of nano or
micron size that are easier to tailor for specific functionalities and to observe.
Colloidal crystals, for example, play a vital role in the study of structural
properties of crystalline systems and the development of engineered, functional
materials \cite{Osterman07,Chen11,*Mao13,Mikhael08}.
In addition, another novel technique for artificial lattice ordering is built on
the trapping of ultracold atoms (e.g., $^{87}$Rb) in optical superlattices produced
by overlaying laser beams \cite{Soltan11,Jo12}, as utilized for the study
of many-body quantum physics.

These 2D systems involve a wide variety of constituent particles
with very different types of microscopic interactions, but exhibit similar
crystalline symmetries such as honeycomb (as for graphene
\cite{Geim07}, silicene \cite{Vogt12}, colloidal crystal \cite{Osterman07},
and lattice of ultracold atoms \cite{Soltan11}), kagome (as realized for colloids
\cite{Chen11,*Mao13} and ultracold $^{87}$Rb \cite{Jo12}), and simple Bravais
lattices like triangular and square \cite{Osterman07,Jo12}.
Thus it is of fundamental importance to identify the universal mechanisms
underlying these distinct modes of crystallization, based on the general
principle of symmetry \cite{McTague78}.
It is also important to understand the nature of topological
defects which occur frequently in such systems and are known to
determine the electronic and mechanical properties of the sample
\cite{Pantelides12}.  Unfortunately it is very difficult to model and predict
the nature of such defected states, due to multiple length and time scales
involved in the non-equilibrium crystallization processes.

In this work we develop a dynamic model that can be applied to the study of
crystallization with a variety of ordered and defected structures. We adopt the
phase field crystal (PFC) formalism
\cite{re:elder02,*re:elder04,re:elder07,Emmerich12,Provatas10},
in the spirit of the Alexander-McTague analysis of crystallization based on Landau
theory \cite{McTague78}. The advantage of this PFC approach is that one can study
polycrystal formation in terms of the atomic number density on diffusive time scales
that are many orders of magnitude larger than that of classical microscopic models
such as molecular dynamics. One can also apply renormalization techniques
\cite{re:goldenfeld05,re:huang10b,Huang13} on the PFC equation to study problems that
involve both micro and meso scales such as epitaxial growth \cite{re:huang08,*re:huang10}
and surface patterning in ultra-thin films \cite{Elder12}.

Recently a great deal of progress has been made on generalizing the PFC
formulation to include more crystal symmetries \cite{kh,re:greenwood10,Wu10,Wu10b},
although in 2D current PFC studies are restricted to triangular and square states.
The basic idea is to incorporate interparticle interactions through a
two-point direct correlation function that (i) has $N$ peaks in Fourier
space (corresponding to $N$ different characteristic length scales) and
(ii) is isotropic.
This allows one to systematically interpolate between different crystalline
states without \textit{a priori} assumptions about any orientation-dependent
interactions and thus allows the study of polycrystalline materials.  Here
we exploit this idea and show that systems with three modes (i.e., $N=3$)
exhibit a surprisingly rich phase behavior of crystallization that covers
symmetries of all five 2D Bravais lattices.
Our results add to a growing list of structures that
can be realized from the freezing of monatomic fluids with isotropic multi-well
interaction potentials \cite{Rechtsman05,Engel07,Edlund11,*Edlund12,Batten11},
and more importantly, provide a systematic approach for examining both structural
and dynamic properties of 2D crystalline materials.

The multi-mode phase field crystal model we introduce here is based on a
dimensionless free energy functional
\begin{eqnarray}
& \mathcal{F} = \int d\vec{r} &
\left \{ \frac{\psi}{2} \left ( r+\lambda\prod_{i=0}^{N-1}[(Q_i^2+\nabla^2)^2+b_i]
\right ) \psi \right. \nonumber\\
&& \left. - \frac{\tau}{3} \psi^3 +\frac{\psi^4}{4} \right \}, \label{Eq.energy}
\end{eqnarray}
as generalized from the two-mode form proposed before \cite{Lifshitz97,Wu10},
and a dynamic equation $\partial \psi / \partial t= \nabla^2
\delta \mathcal{F} / \delta \psi$ on diffusive time scale, giving
\begin{equation}
\partial \psi / \partial t =  \nabla^2 \{ ( r
+\lambda\prod_{i=0}^{N-1}[(Q_i^2+\nabla^2)^2+b_i] ) \psi - \tau \psi^2
+ \psi^3 \},
\label{Eq.model}
\end{equation}
where $\psi(\vec{r},t)$ is a rescaled particle number density field,
and $r$, $\lambda$, $b_i$, and $\tau$ are phenomenological constants.
The parameters $b_i$ control the relative stability of different modes,
and are determined by interparticle potential of a specific system.
This PFC free energy functional can be approximately derived from a
Landau-Brazovskii expansion of the free energy in classical Density Functional
Theory (CDFT) of freezing~\cite{re:elder07,re:huang10b}, and the gradient terms in
Eq. (\ref{Eq.energy}) can be obtained from expanding the Fourier component of
the pair correlation function in CDFT, which satisfies the (i) and (ii) requirements
given above, up to its $N$ peaks that are located at wave numbers $Q_i$
($i=0, 1, ..., N-1$).

In a crystalline state $\psi$ can be expanded in terms
of its Fourier components $A_{\vec{q}}$ and the reciprocal lattice vectors (RLVs)
$\vec{q}$:
$\psi(\vec{r}) =\psi_0+\sum_{\vec{q}} A_{\vec{q}} e^{i\vec{q}\cdot\vec{r}}$,
where $\psi_0$ is the average rescaled density. In 2D,
$\vec{q} = n\vec{k}_1 +m\vec{k}_2$
where $m$ and $n$ are integers, and $\vec{k_1}$ and $\vec{k}_2$ are the
principal RLVs. From Eq.~(\ref{Eq.energy}) we can obtain a standard
expansion form
\begin{eqnarray}
\mathcal{F}/V &=& \sum_{\vec{q}} G_{\vec{q}} A_{\vec{q}} A_{-\vec{q}}
-w\sum_{\vec{q}_1 \vec{q}_2 \vec{q}_3}A_{\vec{q}_1}A_{\vec{q}_2}
A_{\vec{q}_3}\delta_{\vec{q}_1+\vec{q}_2+\vec{q}_3,0} \nonumber\\
&+& \frac{1}{4}
\sum_{\vec{q}_1 \vec{q}_2 \vec{q}_3 \vec{q}_4}A_{\vec{q}_1}
A_{\vec{q}_2}A_{\vec{q}_3}A_{\vec{q}_4}\delta_{\vec{q}_1+\vec{q}_2+\vec{q}_3+\vec{q}_4,0},
\label{Eq.enespec}
\end{eqnarray}
where $V$ is the system volume, $w = \tau/3-\psi_0$, and
\begin{equation}
G_{\vec{q}} = \frac{1}{2} \left \{ r+\lambda\prod_{i=0}^{N-1}[(Q_i^2-q^2)^2+b_i]
\right \} -\tau \psi_0 +\frac{3}{2} \psi_0^2.
\end{equation}
When $G_{\vec{q}}$ is small but negative, a crystalline state forms and the
summation over cubic and quartic terms can be restricted to wavevectors with
magnitude $|\vec{q}|=Q_i$, with higher order harmonics not needed.  It was
noted by Alexander and McTague \cite{McTague78} that close to the melting
point the favored crystalline state is determined by the largest contribution
of the cubic term which, according to Eq. (\ref{Eq.enespec}), is given
by a triplet of density waves with wavevectors forming a closed loop, i.e.,
$\vec{q}_1+\vec{q}_2+\vec{q}_3=0$.

Within the five 2D Bravais lattices the least symmetric one is oblique, a chiral
lattice, for which the triplet of the density waves must consist of wavevectors
with different magnitudes forming a scalene triangular loop. Thus it
is a candidate of preferred state for Eq.~(\ref{Eq.enespec}) when $N = 3$.
The same argument holds for the rectangular lattice. The square or rhombic
lattice can be considered as special cases of the rectangular or oblique lattice
that are stabilized with $N =2$ and with triads of wavevectors forming an
isosceles triangular loop. This $N =2$ limit was explored by Lifshitz and
Petrich \cite{Lifshitz97}, showing stable patterns of $2$-, $4$-, $6$-, and
$12$-fold symmetries. The $N =1$ limit, with the basic wavevectors forming an
equilateral triangular, corresponds to the favored 2D triangular phase as
given in the classical work of Alexander and McTague \cite{McTague78}.
Thus three modes (with different $Q_i$, $i=0,1,2$) are enough for
constructing a minimal model to cover all five 2D Bravais lattices.
Furthermore, the selection and competition between these modes of
different length scales will lead to much richer crystalline phases,
an effect that goes beyond the classical
Alexander-McTague type analysis. As shown below, we can tune the
excitation level of the density waves of $|\vec{q}|= Q_i$ via parameters
$b_i$ in our PFC model to systematically explore the stability of different
phases that compete with a targeted crystalline state.

\begin{figure}
\centerline{\includegraphics[width=0.5\textwidth]{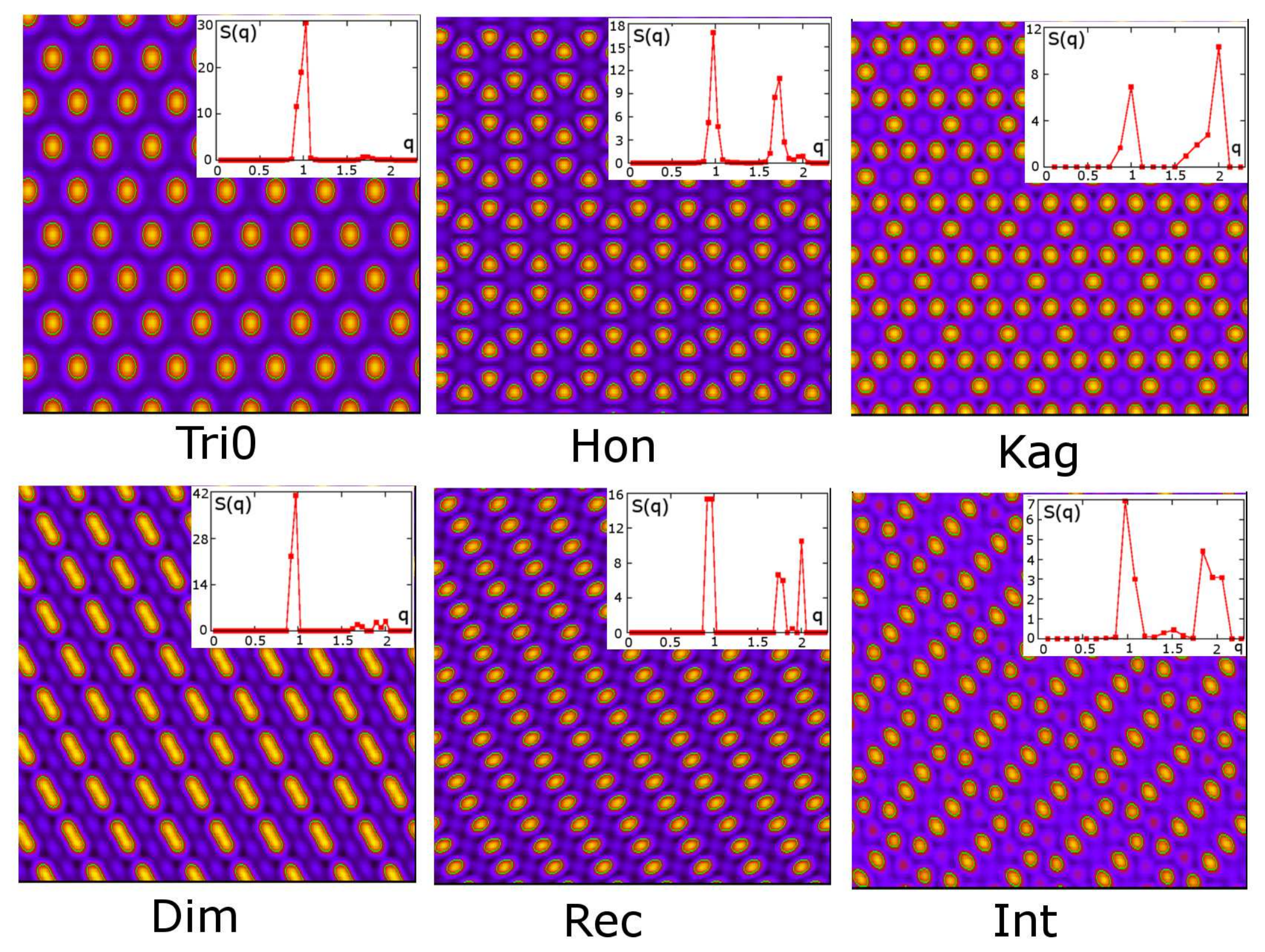}}
\caption{(Color online) Crystalline phases obtained via PFC simulations for
$Q_{i=0,1,2} = {1,\sqrt{3},2}$, including one of three triangular phases
(Tri0), kagome (Kag), honeycomb (Hon), dimer (Dim), rectangular (Rec),
and also an intermediate phase (Int). Insets: The circularly averaged
structure factor $S(q)$ vs. $q$.} \label{fig1}
\end{figure}

To verify our analysis we solved the PFC dynamic equation (\ref{Eq.model}) with
$N =3$ via a pseudo-spectral algorithm~\cite{re:cheng08,Wu08}, using periodic
boundary conditions in systems of sizes ranging from $256^2$ to $1024^2$.
We restricted our parameter space to $\psi_0 =-0.2$, $r = -0.15$,
$\lambda =0.02$, and $\tau=0$ for simplicity.
To systematically determine the various steady states
we chose $Q_i$ such that the magnitudes of the critical
wavevectors correspond to the three shortest wavenumbers of a targeted
lattice. For 2D Bravais lattice they are given by
\begin{equation}
|\vec{q}| = k_1\sqrt{m^2+\mu^2 n^2+2mn\mu \cos \theta},
\label{Eq.wavenumber}
\end{equation}
where 
$\mu =k_2/k_1$ and $\theta$ is the angle between $\vec{k}_1$ and $\vec{k}_2$.

\begin{figure}
\centerline{\includegraphics[width=0.5\textwidth]{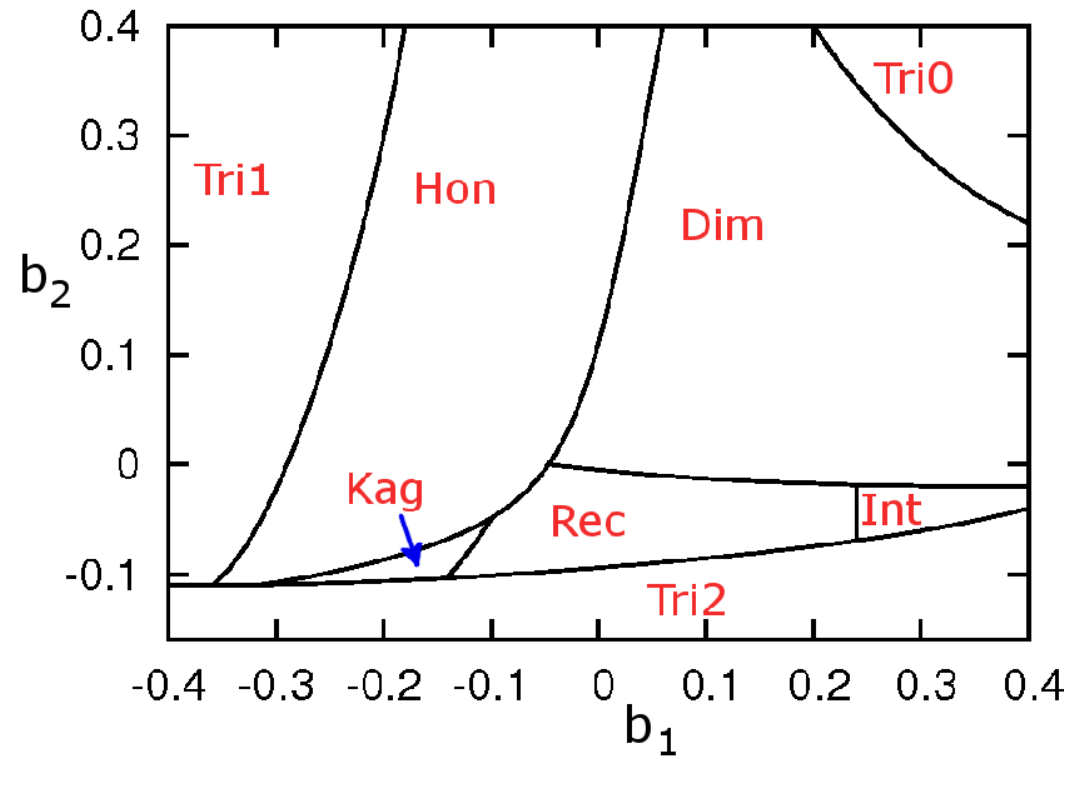}}
\caption{(Color online) Phases determined from the PFC model with
$Q_{i=0,1,2} = {1,\sqrt{3},2}$ and $b_0 =0$.
}
\label{fig2}
\end{figure}

Steady-state solutions were obtained by monitoring the crystallization
process until changes in the system free energy density are negligible
(e.g., $\Delta f < 0.01\%$).
In Fig.~\ref{fig1} we show a variety of ordered states obtained when
$Q_{i=0,1,2} = {1,\,\sqrt{3},\,2}$ (corresponding to the first three shortest
RLVs for triangular lattice), at different regions of the $b_i$ parameter
space. They include: three triangular (Tri0, Tri1, Tri2), honeycomb (Hon),
kagome (Kag), rectangular (Rec), dimer (Dim), and intermediate (Int) phases.
We identified the observed regions of these different states by re-running the
simulations (using different random initial conditions) at each point of the
parameter space for more than 10 times and classifying the stable structure as
the equilibrium phase. The results are depicted in Fig.~\ref{fig2}.

\begin{figure}
\centerline{\includegraphics[width=0.5\textwidth]{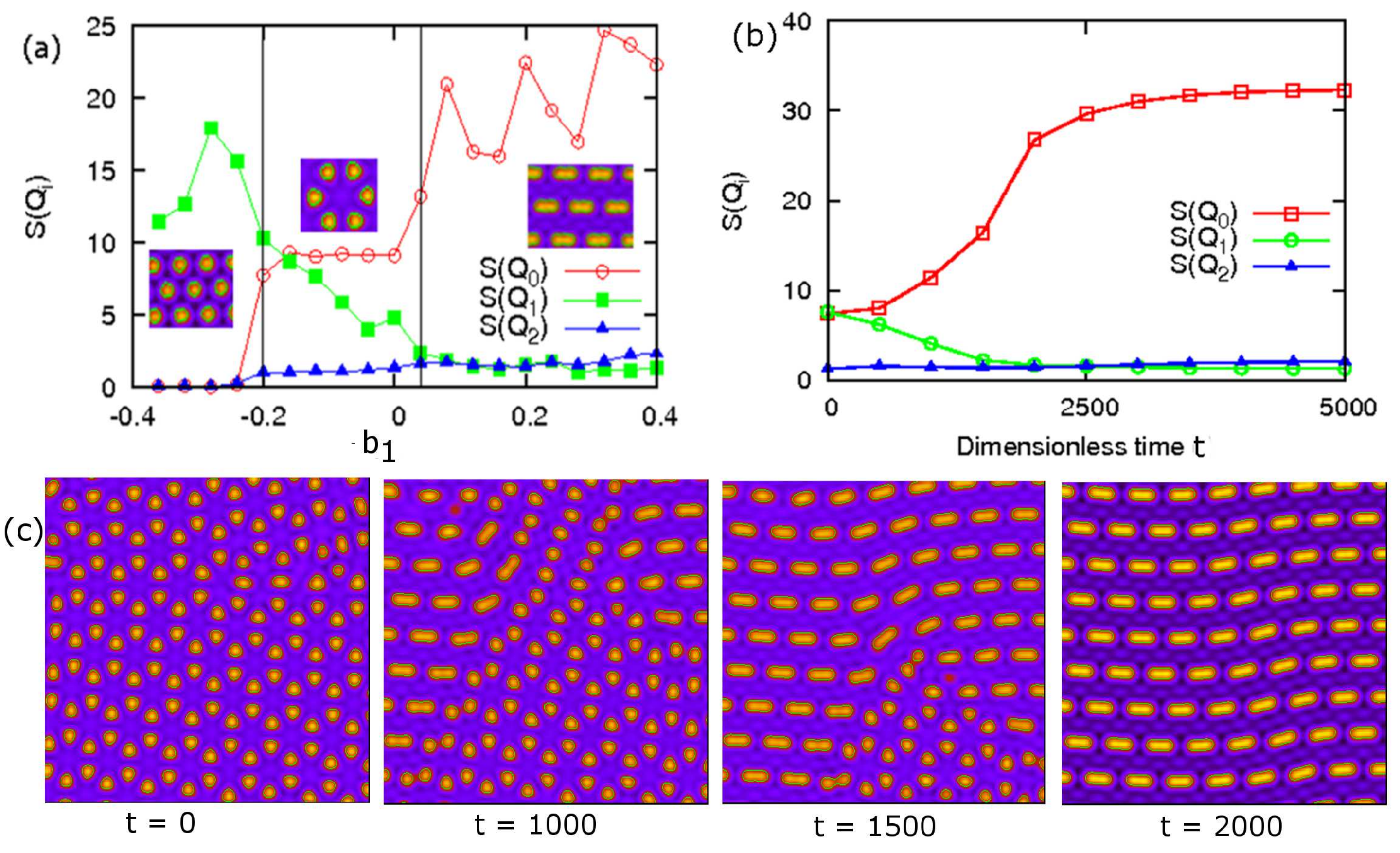}}
\caption{(Color online) Dynamic process of phase transformation. (a) Peak values of
structure factor during a Tri1 $\rightarrow$ Hon $\rightarrow$ Dim transformation,
for $(b_0, b_2)=(0,0.1)$.
(b) and (c) show the dynamics of Hon $\rightarrow$ Dim transformation
from $(b_0, b_1, b_2)=(0, -0.1, 0.04)$ to $(0, 0.2, 0.04)$.
} \label{fig3}
\end{figure}

These simulation results are consistent with the above crystallization analysis.
The stable triangular states are characterized by a circularly averaged
structure factor $S(q)$ with one dominate peak, as shown in Fig.~\ref{fig1}.
The honeycomb phase corresponds to a superposition of two sets of
triplet density waves with $|\vec{q}|$ = $Q_0$ and $Q_1$, respectively. Each
set can maximize the cubic free energy term since the wave vectors can form a
close loop (equilateral triangle). Similar arguments can be made for the kagome
phase, but with each set having wave vectors $|\vec{q}|$ = $Q_0$ and $Q_2$,
respectively. This has been demonstrated in the experiments of ultracold
atoms \cite{Jo12}, where two sets of three optical waves with
$|\vec{q}|$ = $Q_0$ and $Q_2=2Q_0$ were superimposed to create a kagome lattice.
To further examine the formation condition of honeycomb phase we analyze the
following transformation: Tri1 $\rightarrow$ Hon $\rightarrow$ Dim.
The Tri1 $\rightarrow$ Hon transformation is characterized by a sudden increase
of the structure-factor peak at $Q_0$, leading to two prominent peaks in the
honeycomb phase [see Fig.~\ref{fig3}(a)]. A further increase in $b_1$
creates an imbalances between the two sets of critical modes, inducing a
compressed-honeycomb, i.e., dimer state. Fig.~\ref{fig3} shows
the dynamics of the Hon $\rightarrow$ Dim transformation. A pair of density
maxima merge to form elongated regions of higher densities (i.e., dimers)
during the transition.

As discussed above, three modes are needed to form a rectangular phase,
which is verified in our results of Fig.~\ref{fig1}.
Our numerical results also reveal that
one can interpolate between the two Bravais lattice symmetries, triangular and
rectangular, by tuning the excitation levels of the dominant density waves via
$b_i$. This is not surprising since from Eq.~(\ref{Eq.wavenumber}) one can see
that the magnitudes of RLVs in a triangular lattice, $Q_{i=0,1,2} =
{1, \sqrt{3}, 2}$, are the same as those of a rectangular lattice with
$\mu =\sqrt{3}$.
We have realized a stable intermediate state between these two lattices
(see Fig.~\ref{fig1}), which consists of rectangular domains separated
periodically by triangular edges. An analogous phase has been observed in
experiments involving commensurate phase ordering of colloid monolayers
(i.e., archimedean-like tiling \cite{Mikhael08}).

\begin{figure}
\centerline{\includegraphics[width=0.5\textwidth]{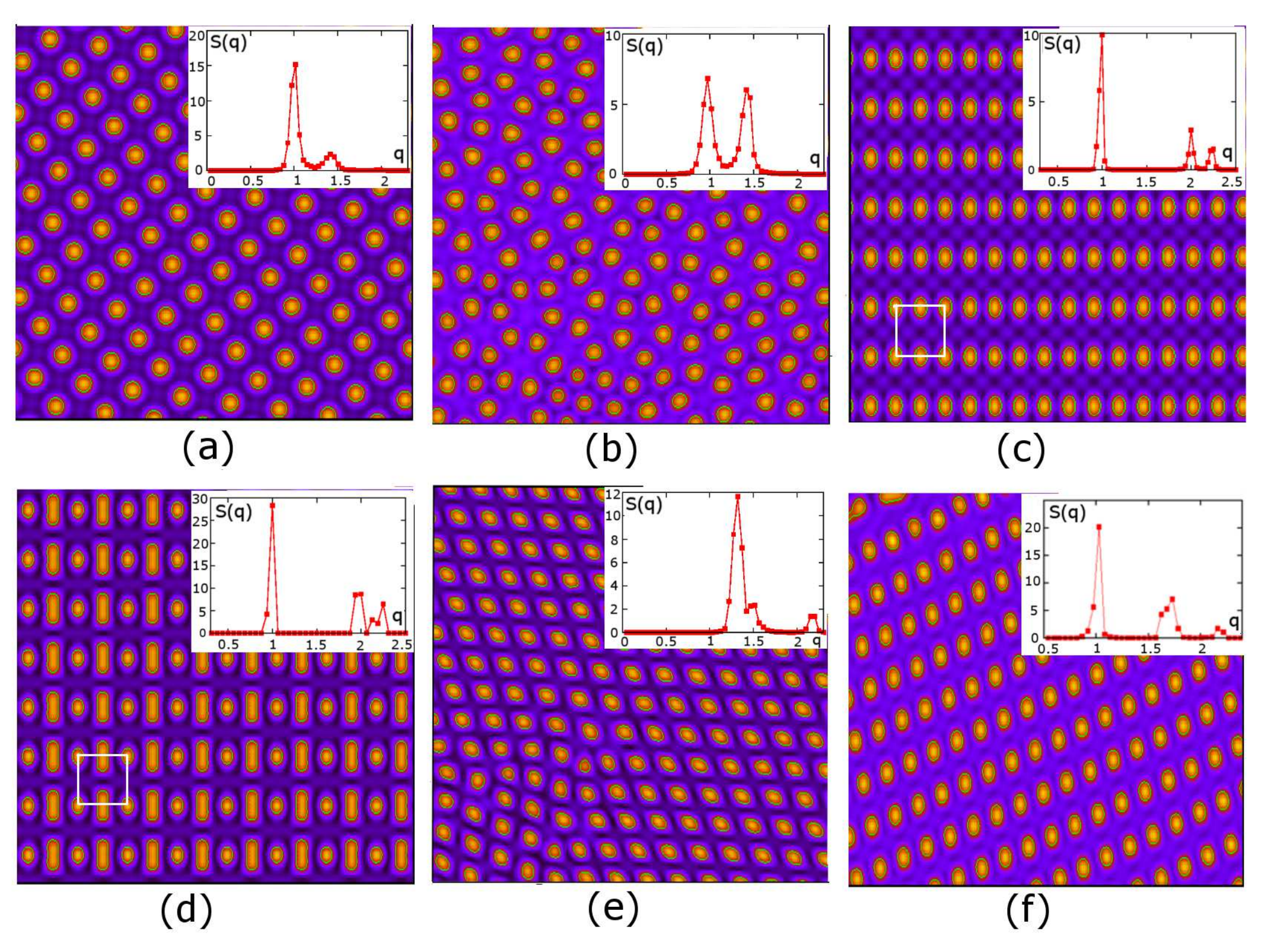}}
\caption{(Color online)
(a) Square and (b) pentagon-hexagon structures are obtained with
$Q_{i=0,1,2} = {1, \sqrt{2}, 2}$.
Example phases with $Q_{i=0,1,2} = {1,\,2,\,\sqrt{5}}$
are shown as (c) a rectangular phase and (d) a dimer-square
crystal, with their square unit cells indicated.
Also (e) a rhombic phase at $Q_{i=0,1,2} = 1.35, 1.57, 2.2$
and (f) an oblique phase at $Q_{i=0,1,2} = 1, \sqrt{3}, 2.2$ are given.
}
\label{fig4}
\end{figure}

We can also target the ordering into
other 2D Bravais lattices: square, rhombic and oblique, by applying
Eq.~(\ref{Eq.wavenumber}). The ratio of the three shortest wavevectors in a
square lattice is given by $Q_{i=0,1,2} = {1,\,\sqrt{2},\,2}$, leading to two
possible sets of density wave triplet: two with $|\vec{q}|=1$ and one
with $|\vec{q}| =\sqrt{2}$, or two with $|\vec{q}|=\sqrt{2}$
and one with $|\vec{q}| = 2$. Both have been obtained in our
simulations, with an example shown in Fig.~\ref{fig4}(a).
We also observed other stable states with the same $Q_i$ series, including
three triangular states and a phase that consists of pentagons
and hexagons [Fig. \ref{fig4}(b)] with dominant
structure-factor peaks located at $|\vec{q}|=Q_0$ and $Q_1$.
A similar pentagon phase was found in recent molecular dynamics simulations
using a double-well potential \cite{Engel07}.

The square-type states can be also generated from the series
$Q_{i=0,1,2} = 1, 2, \sqrt{5}$, although with multi-atom basis. Note that
this $Q_i$ series corresponds to a close loop of right-angled-triangle
wavevectors, and thus a rectangular state, as seen in Fig.~\ref{fig4}(c).
However, such phase can be defined as a square lattice with a
two-atom basis since this $Q_{i}$ series also incorporates the ratio of the
Bragg peak positions of a square structure. Another stable multi-atom square
state, a square-dimer phase, is shown in Fig.~\ref{fig4}(d).

To reproduce a rhombic or oblique phase, we
note that in general the oblique state is favored by the cubic term of free
energy expansion when $|b_i| \ll 1$, where all the three critical modes are
excited, while the rhombic phase is favored when two different modes are
dominant. As shown in our results of Figs.~\ref{fig4}(e) and \ref{fig4}(f),
the structure factor of oblique phase is characterized by three peaks and
the rhombic phase by two dominant peaks, as expected.

\begin{figure}
\centerline{\includegraphics[width=0.5\textwidth]{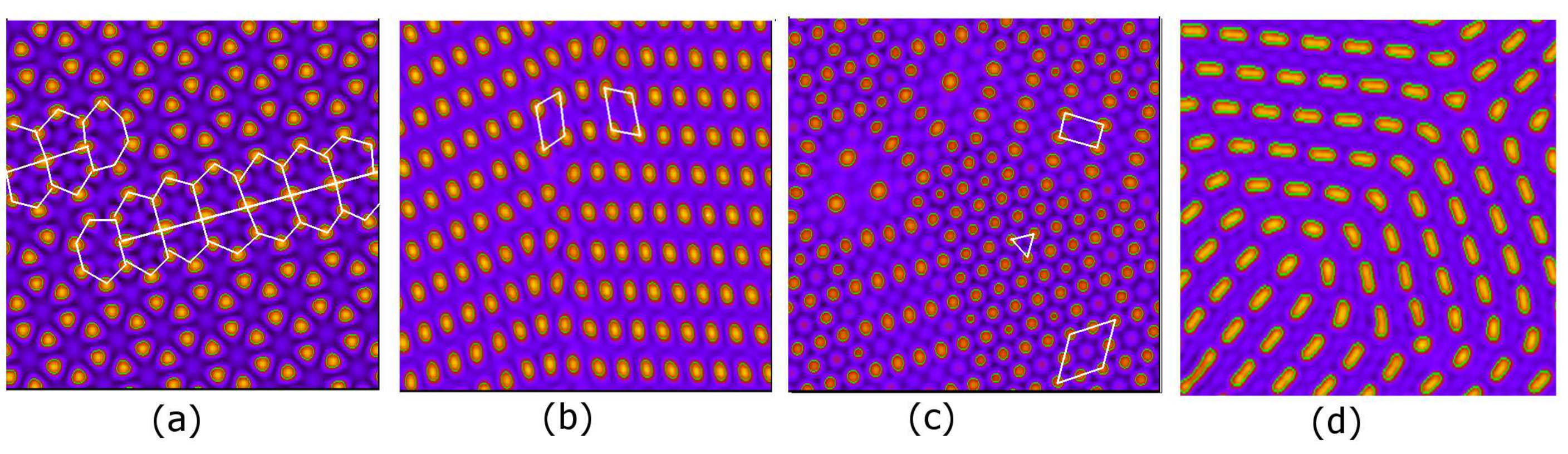}}
\caption{(Color online) Snapshots of defect configurations during dynamic
simulations, for (a) honeycomb phase, (b) oblique phase, (c) coexisting
phases, and (d) dimer phase.}
\label{fig5}
\end{figure}

Another focus of our work is on exploring the dynamics of different crystalline
states that are generated by the three-mode PFC model. Starting from an
unstable, homogeneous liquid state, the typical crystallization process of
a large system involves the nucleation of crystal seeds, the
formation and later the annihilation of topological defects which leads to
the growth and coarsening of crystal grains.  In Fig.~\ref{fig5} we show a
variety of topological defects observed during the ordering process.
This includes linear and point defects in the honeycomb phase, grain
boundaries in the oblique lattice, complex defected state associated with
coexistence of different crystalline structures,
and also disclinations in the dimer state.

The simplicity of our approach also makes it relatively straightforward
to calculate the elastic properties of these crystalline states.
As shown in Fig.~\ref{fig1}, the structure factor of
dimer phase has a dominant single peak and thus a one-mode approximation
\cite{re:elder04,Provatas10,Mkhonta13}
can be utilized. Note also that the dimer structure results from the merging of
two density peaks in a honeycomb phase, as demonstrated in Fig.~\ref{fig3}. Hence
here we can consider $\psi(\vec{r})$ as a two-particle (dimer) density, with
each constituent dimer molecule (basis) consisting of two atoms, one at the
origin and the other at $(\kappa d,\,0)$, where $0<\kappa<1/2$ and $d$ is the
lattice constant of the corresponding triangular lattice they occupy.
Following the standard procedure \cite{Provatas10}, we obtain the shear modulus of
the system $\mu_s = C_{44} = 3\alpha_t[\cos(2\kappa\pi) +1]/8$ where
$\alpha_t = (b_1+4q^4)(b_2+9q^4)\lambda A^2q^4$, $A$ is the amplitude of $\psi$
expansion, and $q =4\pi/(\sqrt{3}d)=Q_0$, and
the anisotropic Poisson ratio parallel and perpendicular to the dimer molecular axis:
\begin{equation}
\nu_{x} =\frac{C_{12}}{C_{22}}= 3\left[ \frac{\cos(2\kappa\pi)+1}
{\cos(2\kappa\pi)+17}\right]\,\, \text{and}\,\,
\nu_{y} =\frac{C_{12}}{C_{11}}=\frac{1}{3}.
\label{Eq.nu}
\end{equation}

When $\kappa \rightarrow 0$ the dimer state changes to a triangular
one (Tri0). In this limit the Poisson ratio becomes isotropic as obtained from
Eq. (\ref{Eq.nu}), which is expected for an elastically isotropic triangular
lattice. Our calculations also show that the shear modulus of the dimer state
is smaller than that of the triangular phase.  This is a consequence of the
additional degree of freedom in this state, i.e., the dimers can rotate
\cite{Mkhonta13}.  When $\kappa \rightarrow 1/2$, the simple one-mode
approximation used here breaks down and more harmonics (modes) are required.
This would correspond to the instability of the dimer phase towards the
formation of the honeycomb phase (which is described by two modes) around the
point of $\kappa = 1/2$.

All our results presented above show that crystallization is not only a
general problem of symmetry as was first argued by Alexander and McTague three
decades ago \cite{McTague78}, but also a problem involving competition and
coupling between different length scales of the system.
As demonstrated, three modes are
enough to produce all 5 Bravais lattices in 2D as well as many of the
non-Bravais structures, including honeycomb and kagome phases that have been
found in novel 2D crystalline materials, and also predictions of more complex
phases. The minimal model presented here can be exploited to study not only
the non-equilibrium formation of crystals and polycrystals with a large
variety of crystalline symmetries, but also the elastic and plastic properties
of such systems \cite{re:tegze11,re:berry08,Chan10}.
Our results can also serve as a guide to experiments on producing or self-assembling
a variety of ordered phases that can form in systems with competing multiple
scales, such as the ordering process of surface-functionalized colloidal
particles or of ultracold atoms in tunable commensurate optical lattices.
The study of such self-assembly process and the evolution of defected
state requires a dynamic modeling method at time scales of experimental relevance,
for which the multi-mode PFC model described here is much more applicable than
conventional atomistic techniques. Furthermore, our modeling framework
can be readily extended to a systematic study of three-dimensional crystalline
and polycrystalline materials or self-assembled systems.

\begin{acknowledgments}
We acknowledge support from the National Science Foundation under Grant No.
DMR-0845264 (Z.-F.H.) and DMR-0906676 (K.R.E.).
\end{acknowledgments}

\bibliography{../references}

\begin{thebibliography}{40}%
\makeatletter
\providecommand \@ifxundefined [1]{%
 \@ifx{#1\undefined}
}%
\providecommand \@ifnum [1]{%
 \ifnum #1\expandafter \@firstoftwo
 \else \expandafter \@secondoftwo
 \fi
}%
\providecommand \@ifx [1]{%
 \ifx #1\expandafter \@firstoftwo
 \else \expandafter \@secondoftwo
 \fi
}%
\providecommand \natexlab [1]{#1}%
\providecommand \enquote  [1]{``#1''}%
\providecommand \bibnamefont  [1]{#1}%
\providecommand \bibfnamefont [1]{#1}%
\providecommand \citenamefont [1]{#1}%
\providecommand \href@noop [0]{\@secondoftwo}%
\providecommand \href [0]{\begingroup \@sanitize@url \@href}%
\providecommand \@href[1]{\@@startlink{#1}\@@href}%
\providecommand \@@href[1]{\endgroup#1\@@endlink}%
\providecommand \@sanitize@url [0]{\catcode `\\12\catcode `\$12\catcode
  `\&12\catcode `\#12\catcode `\^12\catcode `\_12\catcode `\%12\relax}%
\providecommand \@@startlink[1]{}%
\providecommand \@@endlink[0]{}%
\providecommand \url  [0]{\begingroup\@sanitize@url \@url }%
\providecommand \@url [1]{\endgroup\@href {#1}{\urlprefix }}%
\providecommand \urlprefix  [0]{URL }%
\providecommand \Eprint [0]{\href }%
\providecommand \doibase [0]{http://dx.doi.org/}%
\providecommand \selectlanguage [0]{\@gobble}%
\providecommand \bibinfo  [0]{\@secondoftwo}%
\providecommand \bibfield  [0]{\@secondoftwo}%
\providecommand \translation [1]{[#1]}%
\providecommand \BibitemOpen [0]{}%
\providecommand \bibitemStop [0]{}%
\providecommand \bibitemNoStop [0]{.\EOS\space}%
\providecommand \EOS [0]{\spacefactor3000\relax}%
\providecommand \BibitemShut  [1]{\csname bibitem#1\endcsname}%
\let\auto@bib@innerbib\@empty
\bibitem [{\citenamefont {Geim}\ and\ \citenamefont
  {Novoselov}(2007)}]{Geim07}%
  \BibitemOpen
  \bibfield  {author} {\bibinfo {author} {\bibfnamefont {A.~K.}\ \bibnamefont
  {Geim}}\ and\ \bibinfo {author} {\bibfnamefont {K.~S.}\ \bibnamefont
  {Novoselov}},\ }\href@noop {} {\bibfield  {journal} {\bibinfo  {journal}
  {Nature Mater.}\ }\textbf {\bibinfo {volume} {6}},\ \bibinfo {pages} {183}
  (\bibinfo {year} {2007})}\BibitemShut {NoStop}%
\bibitem [{\citenamefont {Pantelides}\ \emph {et~al.}(2012)\citenamefont
  {Pantelides}, \citenamefont {Puzyrev}, \citenamefont {Tsetseris},\ and\
  \citenamefont {Wang}}]{Pantelides12}%
  \BibitemOpen
  \bibfield  {author} {\bibinfo {author} {\bibfnamefont {S.~T.}\ \bibnamefont
  {Pantelides}}, \bibinfo {author} {\bibfnamefont {Y.}~\bibnamefont {Puzyrev}},
  \bibinfo {author} {\bibfnamefont {L.}~\bibnamefont {Tsetseris}}, \ and\
  \bibinfo {author} {\bibfnamefont {B.}~\bibnamefont {Wang}},\ }\href@noop {}
  {\bibfield  {journal} {\bibinfo  {journal} {MRS Bulletin}\ }\textbf {\bibinfo
  {volume} {37}},\ \bibinfo {pages} {1187} (\bibinfo {year}
  {2012})}\BibitemShut {NoStop}%
\bibitem [{\citenamefont {Vogt}\ \emph {et~al.}(2012)\citenamefont {Vogt},
  \citenamefont {De~Padova}, \citenamefont {Quaresima}, \citenamefont {Avila},
  \citenamefont {Frantzeskakis}, \citenamefont {Asensio}, \citenamefont
  {Resta}, \citenamefont {Ealet},\ and\ \citenamefont {Le~Lay}}]{Vogt12}%
  \BibitemOpen
  \bibfield  {author} {\bibinfo {author} {\bibfnamefont {P.}~\bibnamefont
  {Vogt}}, \bibinfo {author} {\bibfnamefont {P.}~\bibnamefont {De~Padova}},
  \bibinfo {author} {\bibfnamefont {C.}~\bibnamefont {Quaresima}}, \bibinfo
  {author} {\bibfnamefont {J.}~\bibnamefont {Avila}}, \bibinfo {author}
  {\bibfnamefont {E.}~\bibnamefont {Frantzeskakis}}, \bibinfo {author}
  {\bibfnamefont {M.~C.}\ \bibnamefont {Asensio}}, \bibinfo {author}
  {\bibfnamefont {A.}~\bibnamefont {Resta}}, \bibinfo {author} {\bibfnamefont
  {B.}~\bibnamefont {Ealet}}, \ and\ \bibinfo {author} {\bibfnamefont
  {G.}~\bibnamefont {Le~Lay}},\ }\href@noop {} {\bibfield  {journal} {\bibinfo
  {journal} {Phys. Rev. Lett.}\ }\textbf {\bibinfo {volume} {108}},\ \bibinfo
  {pages} {155501} (\bibinfo {year} {2012})}\BibitemShut {NoStop}%
\bibitem [{\citenamefont {Bianco}\ \emph {et~al.}(2013)\citenamefont {Bianco},
  \citenamefont {Butler}, \citenamefont {Jiang}, \citenamefont {Restrepo},
  \citenamefont {Windl},\ and\ \citenamefont {Goldberger}}]{Bianco13}%
  \BibitemOpen
  \bibfield  {author} {\bibinfo {author} {\bibfnamefont {E.}~\bibnamefont
  {Bianco}}, \bibinfo {author} {\bibfnamefont {S.}~\bibnamefont {Butler}},
  \bibinfo {author} {\bibfnamefont {S.}~\bibnamefont {Jiang}}, \bibinfo
  {author} {\bibfnamefont {O.~D.}\ \bibnamefont {Restrepo}}, \bibinfo {author}
  {\bibfnamefont {W.}~\bibnamefont {Windl}}, \ and\ \bibinfo {author}
  {\bibfnamefont {J.~E.}\ \bibnamefont {Goldberger}},\ }\href {\doibase DOI:
  10.1021/nn4009406} {\bibfield  {journal} {\bibinfo  {journal} {ACS Nano}\ }
  (\bibinfo {year} {2013}),\ DOI: 10.1021/nn4009406}\BibitemShut {NoStop}%
\bibitem [{\citenamefont {Ci}\ \emph {et~al.}(2010)\citenamefont {Ci},
  \citenamefont {Song}, \citenamefont {Jin}, \citenamefont {Jariwala},
  \citenamefont {Wu}, \citenamefont {Li}, \citenamefont {Srivastava},
  \citenamefont {Wang}, \citenamefont {Storr}, \citenamefont {Balicas},
  \citenamefont {Liu},\ and\ \citenamefont {Ajayan}}]{Ci10}%
  \BibitemOpen
  \bibfield  {author} {\bibinfo {author} {\bibfnamefont {L.}~\bibnamefont
  {Ci}}, \bibinfo {author} {\bibfnamefont {L.}~\bibnamefont {Song}}, \bibinfo
  {author} {\bibfnamefont {C.}~\bibnamefont {Jin}}, \bibinfo {author}
  {\bibfnamefont {D.}~\bibnamefont {Jariwala}}, \bibinfo {author}
  {\bibfnamefont {D.}~\bibnamefont {Wu}}, \bibinfo {author} {\bibfnamefont
  {Y.}~\bibnamefont {Li}}, \bibinfo {author} {\bibfnamefont {A.}~\bibnamefont
  {Srivastava}}, \bibinfo {author} {\bibfnamefont {Z.~F.}\ \bibnamefont
  {Wang}}, \bibinfo {author} {\bibfnamefont {K.}~\bibnamefont {Storr}},
  \bibinfo {author} {\bibfnamefont {L.}~\bibnamefont {Balicas}}, \bibinfo
  {author} {\bibfnamefont {F.}~\bibnamefont {Liu}}, \ and\ \bibinfo {author}
  {\bibfnamefont {P.~M.}\ \bibnamefont {Ajayan}},\ }\href@noop {} {\bibfield
  {journal} {\bibinfo  {journal} {Nature Mater.}\ }\textbf {\bibinfo {volume}
  {9}},\ \bibinfo {pages} {430} (\bibinfo {year} {2010})}\BibitemShut {NoStop}%
\bibitem [{\citenamefont {Tongay}\ \emph {et~al.}(2012)\citenamefont {Tongay},
  \citenamefont {Zhou}, \citenamefont {Ataca}, \citenamefont {Lo},
  \citenamefont {Matthews}, \citenamefont {Li}, \citenamefont {Grossman},\ and\
  \citenamefont {Wu}}]{Tongay12}%
  \BibitemOpen
  \bibfield  {author} {\bibinfo {author} {\bibfnamefont {S.}~\bibnamefont
  {Tongay}}, \bibinfo {author} {\bibfnamefont {J.}~\bibnamefont {Zhou}},
  \bibinfo {author} {\bibfnamefont {C.}~\bibnamefont {Ataca}}, \bibinfo
  {author} {\bibfnamefont {K.}~\bibnamefont {Lo}}, \bibinfo {author}
  {\bibfnamefont {T.~S.}\ \bibnamefont {Matthews}}, \bibinfo {author}
  {\bibfnamefont {J.}~\bibnamefont {Li}}, \bibinfo {author} {\bibfnamefont
  {J.~C.}\ \bibnamefont {Grossman}}, \ and\ \bibinfo {author} {\bibfnamefont
  {J.}~\bibnamefont {Wu}},\ }\href@noop {} {\bibfield  {journal} {\bibinfo
  {journal} {Nano Lett.}\ }\textbf {\bibinfo {volume} {12}},\ \bibinfo {pages}
  {5576} (\bibinfo {year} {2012})}\BibitemShut {NoStop}%
\bibitem [{\citenamefont {Osterman}\ \emph {et~al.}(2007)\citenamefont
  {Osterman}, \citenamefont {Babic}, \citenamefont {Poberaj}, \citenamefont
  {Dobnikar},\ and\ \citenamefont {Ziherl}}]{Osterman07}%
  \BibitemOpen
  \bibfield  {author} {\bibinfo {author} {\bibfnamefont {N.}~\bibnamefont
  {Osterman}}, \bibinfo {author} {\bibfnamefont {D.}~\bibnamefont {Babic}},
  \bibinfo {author} {\bibfnamefont {I.}~\bibnamefont {Poberaj}}, \bibinfo
  {author} {\bibfnamefont {J.}~\bibnamefont {Dobnikar}}, \ and\ \bibinfo
  {author} {\bibfnamefont {P.}~\bibnamefont {Ziherl}},\ }\href@noop {}
  {\bibfield  {journal} {\bibinfo  {journal} {Phys. Rev. Lett.}\ }\textbf
  {\bibinfo {volume} {99}},\ \bibinfo {pages} {248301} (\bibinfo {year}
  {2007})}\BibitemShut {NoStop}%
\bibitem [{\citenamefont {Chen}\ \emph {et~al.}(2011)\citenamefont {Chen},
  \citenamefont {Bae},\ and\ \citenamefont {Granick}}]{Chen11}%
  \BibitemOpen
  \bibfield  {author} {\bibinfo {author} {\bibfnamefont {Q.}~\bibnamefont
  {Chen}}, \bibinfo {author} {\bibfnamefont {S.~C.}\ \bibnamefont {Bae}}, \
  and\ \bibinfo {author} {\bibfnamefont {S.}~\bibnamefont {Granick}},\
  }\href@noop {} {\bibfield  {journal} {\bibinfo  {journal} {Nature}\ }\textbf
  {\bibinfo {volume} {469}},\ \bibinfo {pages} {381} (\bibinfo {year}
  {2011})}\BibitemShut {NoStop}%
\bibitem [{\citenamefont {Mao}\ \emph {et~al.}(2013)\citenamefont {Mao},
  \citenamefont {Chen},\ and\ \citenamefont {S.}}]{Mao13}%
  \BibitemOpen
  \bibfield  {author} {\bibinfo {author} {\bibfnamefont {X.}~\bibnamefont
  {Mao}}, \bibinfo {author} {\bibfnamefont {Q.}~\bibnamefont {Chen}}, \ and\
  \bibinfo {author} {\bibfnamefont {G.}~\bibnamefont {S.}},\ }\href@noop {}
  {\bibfield  {journal} {\bibinfo  {journal} {Nature Mater.}\ }\textbf
  {\bibinfo {volume} {12}},\ \bibinfo {pages} {217} (\bibinfo {year}
  {2013})}\BibitemShut {NoStop}%
\bibitem [{\citenamefont {Mikhael}\ \emph {et~al.}(2008)\citenamefont
  {Mikhael}, \citenamefont {Roth}, \citenamefont {Helden},\ and\ \citenamefont
  {Bechinger}}]{Mikhael08}%
  \BibitemOpen
  \bibfield  {author} {\bibinfo {author} {\bibfnamefont {J.}~\bibnamefont
  {Mikhael}}, \bibinfo {author} {\bibfnamefont {J.}~\bibnamefont {Roth}},
  \bibinfo {author} {\bibfnamefont {L.}~\bibnamefont {Helden}}, \ and\ \bibinfo
  {author} {\bibfnamefont {C.}~\bibnamefont {Bechinger}},\ }\href@noop {}
  {\bibfield  {journal} {\bibinfo  {journal} {Nature}\ }\textbf {\bibinfo
  {volume} {454}},\ \bibinfo {pages} {501} (\bibinfo {year}
  {2008})}\BibitemShut {NoStop}%
\bibitem [{\citenamefont {Soltan-Panahi}\ \emph {et~al.}(2011)\citenamefont
  {Soltan-Panahi}, \citenamefont {Struck}, \citenamefont {Hauke}, \citenamefont
  {Bick}, \citenamefont {Plenkers}, \citenamefont {Meineke}, \citenamefont
  {Becker}, \citenamefont {P.Windpassinger}, \citenamefont {Lewenstein},\ and\
  \citenamefont {Sengstock}}]{Soltan11}%
  \BibitemOpen
  \bibfield  {author} {\bibinfo {author} {\bibfnamefont {P.}~\bibnamefont
  {Soltan-Panahi}}, \bibinfo {author} {\bibfnamefont {J.}~\bibnamefont
  {Struck}}, \bibinfo {author} {\bibfnamefont {P.}~\bibnamefont {Hauke}},
  \bibinfo {author} {\bibfnamefont {A.}~\bibnamefont {Bick}}, \bibinfo {author}
  {\bibfnamefont {W.}~\bibnamefont {Plenkers}}, \bibinfo {author}
  {\bibfnamefont {G.}~\bibnamefont {Meineke}}, \bibinfo {author} {\bibfnamefont
  {C.}~\bibnamefont {Becker}}, \bibinfo {author} {\bibnamefont
  {P.Windpassinger}}, \bibinfo {author} {\bibfnamefont {M.}~\bibnamefont
  {Lewenstein}}, \ and\ \bibinfo {author} {\bibfnamefont {K.}~\bibnamefont
  {Sengstock}},\ }\href@noop {} {\bibfield  {journal} {\bibinfo  {journal}
  {Nature Phys.}\ }\textbf {\bibinfo {volume} {7}} (\bibinfo {year}
  {2011})}\BibitemShut {NoStop}%
\bibitem [{\citenamefont {Jo}\ \emph {et~al.}(2012)\citenamefont {Jo},
  \citenamefont {Guzman}, \citenamefont {Thomas}, \citenamefont {Hosur},
  \citenamefont {Vishwanath},\ and\ \citenamefont {Stamper-Kurn}}]{Jo12}%
  \BibitemOpen
  \bibfield  {author} {\bibinfo {author} {\bibfnamefont {G.-B.}\ \bibnamefont
  {Jo}}, \bibinfo {author} {\bibfnamefont {J.}~\bibnamefont {Guzman}}, \bibinfo
  {author} {\bibfnamefont {C.~K.}\ \bibnamefont {Thomas}}, \bibinfo {author}
  {\bibfnamefont {P.}~\bibnamefont {Hosur}}, \bibinfo {author} {\bibfnamefont
  {A.}~\bibnamefont {Vishwanath}}, \ and\ \bibinfo {author} {\bibfnamefont
  {D.~M.}\ \bibnamefont {Stamper-Kurn}},\ }\href@noop {} {\bibfield  {journal}
  {\bibinfo  {journal} {Phys. Rev. Lett.}\ }\textbf {\bibinfo {volume} {108}},\
  \bibinfo {pages} {045305} (\bibinfo {year} {2012})}\BibitemShut {NoStop}%
\bibitem [{\citenamefont {Alexander}\ and\ \citenamefont
  {McTague}(1978)}]{McTague78}%
  \BibitemOpen
  \bibfield  {author} {\bibinfo {author} {\bibfnamefont {S.}~\bibnamefont
  {Alexander}}\ and\ \bibinfo {author} {\bibfnamefont {J.}~\bibnamefont
  {McTague}},\ }\href@noop {} {\bibfield  {journal} {\bibinfo  {journal} {Phys.
  Rev. Lett.}\ }\textbf {\bibinfo {volume} {41}},\ \bibinfo {pages} {702}
  (\bibinfo {year} {1978})}\BibitemShut {NoStop}%
\bibitem [{\citenamefont {Elder}\ \emph {et~al.}(2002)\citenamefont {Elder},
  \citenamefont {Katakowski}, \citenamefont {Haataja},\ and\ \citenamefont
  {Grant}}]{re:elder02}%
  \BibitemOpen
  \bibfield  {author} {\bibinfo {author} {\bibfnamefont {K.~R.}\ \bibnamefont
  {Elder}}, \bibinfo {author} {\bibfnamefont {M.}~\bibnamefont {Katakowski}},
  \bibinfo {author} {\bibfnamefont {M.}~\bibnamefont {Haataja}}, \ and\
  \bibinfo {author} {\bibfnamefont {M.}~\bibnamefont {Grant}},\ }\href@noop {}
  {\bibfield  {journal} {\bibinfo  {journal} {Phys. Rev. Lett.}\ }\textbf
  {\bibinfo {volume} {88}},\ \bibinfo {pages} {245701} (\bibinfo {year}
  {2002})}\BibitemShut {NoStop}%
\bibitem [{\citenamefont {Elder}\ and\ \citenamefont
  {Grant}(2004)}]{re:elder04}%
  \BibitemOpen
  \bibfield  {author} {\bibinfo {author} {\bibfnamefont {K.~R.}\ \bibnamefont
  {Elder}}\ and\ \bibinfo {author} {\bibfnamefont {M.}~\bibnamefont {Grant}},\
  }\href@noop {} {\bibfield  {journal} {\bibinfo  {journal} {Phys. Rev. E}\
  }\textbf {\bibinfo {volume} {70}},\ \bibinfo {pages} {051605} (\bibinfo
  {year} {2004})}\BibitemShut {NoStop}%
\bibitem [{\citenamefont {Elder}\ \emph {et~al.}(2007)\citenamefont {Elder},
  \citenamefont {Provatas}, \citenamefont {Berry}, \citenamefont {Stefanovic},\
  and\ \citenamefont {Grant}}]{re:elder07}%
  \BibitemOpen
  \bibfield  {author} {\bibinfo {author} {\bibfnamefont {K.~R.}\ \bibnamefont
  {Elder}}, \bibinfo {author} {\bibfnamefont {N.}~\bibnamefont {Provatas}},
  \bibinfo {author} {\bibfnamefont {J.}~\bibnamefont {Berry}}, \bibinfo
  {author} {\bibfnamefont {P.}~\bibnamefont {Stefanovic}}, \ and\ \bibinfo
  {author} {\bibfnamefont {M.}~\bibnamefont {Grant}},\ }\href@noop {}
  {\bibfield  {journal} {\bibinfo  {journal} {Phys. Rev. B}\ }\textbf {\bibinfo
  {volume} {75}},\ \bibinfo {pages} {064107} (\bibinfo {year}
  {2007})}\BibitemShut {NoStop}%
\bibitem [{\citenamefont {Emmerich}\ \emph {et~al.}(2012)\citenamefont
  {Emmerich}, \citenamefont {L{\"{o}}wen}, \citenamefont {Wittkowski},
  \citenamefont {Gruhn}, \citenamefont {T{\'{o}}th}, \citenamefont {Tegze},\
  and\ \citenamefont {Gr{\'{a}}n{\'{a}}sy}}]{Emmerich12}%
  \BibitemOpen
  \bibfield  {author} {\bibinfo {author} {\bibfnamefont {H.}~\bibnamefont
  {Emmerich}}, \bibinfo {author} {\bibfnamefont {H.}~\bibnamefont
  {L{\"{o}}wen}}, \bibinfo {author} {\bibfnamefont {R.}~\bibnamefont
  {Wittkowski}}, \bibinfo {author} {\bibfnamefont {T.}~\bibnamefont {Gruhn}},
  \bibinfo {author} {\bibfnamefont {G.~I.}\ \bibnamefont {T{\'{o}}th}},
  \bibinfo {author} {\bibfnamefont {G.}~\bibnamefont {Tegze}}, \ and\ \bibinfo
  {author} {\bibfnamefont {L.}~\bibnamefont {Gr{\'{a}}n{\'{a}}sy}},\
  }\href@noop {} {\bibfield  {journal} {\bibinfo  {journal} {Adv. Phys.}\
  }\textbf {\bibinfo {volume} {61}},\ \bibinfo {pages} {665} (\bibinfo {year}
  {2012})}\BibitemShut {NoStop}%
\bibitem [{\citenamefont {Provatas}\ and\ \citenamefont
  {Elder}(2010)}]{Provatas10}%
  \BibitemOpen
  \bibfield  {author} {\bibinfo {author} {\bibfnamefont {N.}~\bibnamefont
  {Provatas}}\ and\ \bibinfo {author} {\bibfnamefont {K.~R.}\ \bibnamefont
  {Elder}},\ }\href@noop {} {\emph {\bibinfo {title} {Phase Field Methods in
  Materials Science and Engineering}}}\ (\bibinfo  {publisher} {Wiley-VCH},\
  \bibinfo {address} {Weinheim},\ \bibinfo {year} {2010})\BibitemShut {NoStop}%
\bibitem [{\citenamefont {Goldenfeld}\ \emph {et~al.}(2005)\citenamefont
  {Goldenfeld}, \citenamefont {Athreya},\ and\ \citenamefont
  {Dantzig}}]{re:goldenfeld05}%
  \BibitemOpen
  \bibfield  {author} {\bibinfo {author} {\bibfnamefont {N.}~\bibnamefont
  {Goldenfeld}}, \bibinfo {author} {\bibfnamefont {B.~P.}\ \bibnamefont
  {Athreya}}, \ and\ \bibinfo {author} {\bibfnamefont {J.~A.}\ \bibnamefont
  {Dantzig}},\ }\href@noop {} {\bibfield  {journal} {\bibinfo  {journal} {Phys.
  Rev. E}\ }\textbf {\bibinfo {volume} {72}},\ \bibinfo {pages} {020601(R)}
  (\bibinfo {year} {2005})}\BibitemShut {NoStop}%
\bibitem [{\citenamefont {Huang}\ \emph {et~al.}(2010)\citenamefont {Huang},
  \citenamefont {Elder},\ and\ \citenamefont {Provatas}}]{re:huang10b}%
  \BibitemOpen
  \bibfield  {author} {\bibinfo {author} {\bibfnamefont {Z.-F.}\ \bibnamefont
  {Huang}}, \bibinfo {author} {\bibfnamefont {K.~R.}\ \bibnamefont {Elder}}, \
  and\ \bibinfo {author} {\bibfnamefont {N.}~\bibnamefont {Provatas}},\
  }\href@noop {} {\bibfield  {journal} {\bibinfo  {journal} {Phys. Rev. E}\
  }\textbf {\bibinfo {volume} {82}},\ \bibinfo {pages} {021605} (\bibinfo
  {year} {2010})}\BibitemShut {NoStop}%
\bibitem [{\citenamefont {Huang}(2013)}]{Huang13}%
  \BibitemOpen
  \bibfield  {author} {\bibinfo {author} {\bibfnamefont {Z.-F.}\ \bibnamefont
  {Huang}},\ }\href@noop {} {\bibfield  {journal} {\bibinfo  {journal} {Phys.
  Rev. E}\ }\textbf {\bibinfo {volume} {87}},\ \bibinfo {pages} {012401}
  (\bibinfo {year} {2013})}\BibitemShut {NoStop}%
\bibitem [{\citenamefont {Huang}\ and\ \citenamefont
  {Elder}(2008)}]{re:huang08}%
  \BibitemOpen
  \bibfield  {author} {\bibinfo {author} {\bibfnamefont {Z.-F.}\ \bibnamefont
  {Huang}}\ and\ \bibinfo {author} {\bibfnamefont {K.~R.}\ \bibnamefont
  {Elder}},\ }\href@noop {} {\bibfield  {journal} {\bibinfo  {journal} {Phys.
  Rev. Lett.}\ }\textbf {\bibinfo {volume} {101}},\ \bibinfo {pages} {158701}
  (\bibinfo {year} {2008})}\BibitemShut {NoStop}%
\bibitem [{\citenamefont {Huang}\ and\ \citenamefont
  {Elder}(2010)}]{re:huang10}%
  \BibitemOpen
  \bibfield  {author} {\bibinfo {author} {\bibfnamefont {Z.-F.}\ \bibnamefont
  {Huang}}\ and\ \bibinfo {author} {\bibfnamefont {K.~R.}\ \bibnamefont
  {Elder}},\ }\href@noop {} {\bibfield  {journal} {\bibinfo  {journal} {Phys.
  Rev. B}\ }\textbf {\bibinfo {volume} {81}},\ \bibinfo {pages} {165421}
  (\bibinfo {year} {2010})}\BibitemShut {NoStop}%
\bibitem [{\citenamefont {Elder}\ \emph {et~al.}(2012)\citenamefont {Elder},
  \citenamefont {Rossi}, \citenamefont {Kanerva}, \citenamefont {Sanches},
  \citenamefont {Ying}, \citenamefont {Granato}, \citenamefont {Achim},\ and\
  \citenamefont {Ala-Nissila}}]{Elder12}%
  \BibitemOpen
  \bibfield  {author} {\bibinfo {author} {\bibfnamefont {K.~R.}\ \bibnamefont
  {Elder}}, \bibinfo {author} {\bibfnamefont {G.}~\bibnamefont {Rossi}},
  \bibinfo {author} {\bibfnamefont {P.}~\bibnamefont {Kanerva}}, \bibinfo
  {author} {\bibfnamefont {F.}~\bibnamefont {Sanches}}, \bibinfo {author}
  {\bibfnamefont {S.-C.}\ \bibnamefont {Ying}}, \bibinfo {author}
  {\bibfnamefont {E.}~\bibnamefont {Granato}}, \bibinfo {author} {\bibfnamefont
  {C.~V.}\ \bibnamefont {Achim}}, \ and\ \bibinfo {author} {\bibfnamefont
  {T.}~\bibnamefont {Ala-Nissila}},\ }\href@noop {} {\bibfield  {journal}
  {\bibinfo  {journal} {Phys. Rev. Lett.}\ }\textbf {\bibinfo {volume} {108}},\
  \bibinfo {pages} {226102} (\bibinfo {year} {2012})}\BibitemShut {NoStop}%
\bibitem [{kh()}]{kh}%
  \BibitemOpen
  \href@noop {} {}\bibinfo {note} {A. G. Khachaturyan, private
  communication.}\BibitemShut {Stop}%
\bibitem [{\citenamefont {Greenwood}\ \emph {et~al.}(2010)\citenamefont
  {Greenwood}, \citenamefont {Provatas},\ and\ \citenamefont
  {Rottler}}]{re:greenwood10}%
  \BibitemOpen
  \bibfield  {author} {\bibinfo {author} {\bibfnamefont {M.}~\bibnamefont
  {Greenwood}}, \bibinfo {author} {\bibfnamefont {N.}~\bibnamefont {Provatas}},
  \ and\ \bibinfo {author} {\bibfnamefont {J.}~\bibnamefont {Rottler}},\
  }\href@noop {} {\bibfield  {journal} {\bibinfo  {journal} {Phys. Rev. Lett.}\
  }\textbf {\bibinfo {volume} {105}},\ \bibinfo {pages} {045702} (\bibinfo
  {year} {2010})}\BibitemShut {NoStop}%
\bibitem [{\citenamefont {Wu}\ \emph {et~al.}(2010{\natexlab{a}})\citenamefont
  {Wu}, \citenamefont {Adland},\ and\ \citenamefont {Karma}}]{Wu10}%
  \BibitemOpen
  \bibfield  {author} {\bibinfo {author} {\bibfnamefont {K.-A.}\ \bibnamefont
  {Wu}}, \bibinfo {author} {\bibfnamefont {A.}~\bibnamefont {Adland}}, \ and\
  \bibinfo {author} {\bibfnamefont {A.}~\bibnamefont {Karma}},\ }\href@noop {}
  {\bibfield  {journal} {\bibinfo  {journal} {Phys. Rev. E}\ }\textbf {\bibinfo
  {volume} {81}},\ \bibinfo {pages} {061601} (\bibinfo {year}
  {2010}{\natexlab{a}})}\BibitemShut {NoStop}%
\bibitem [{\citenamefont {Wu}\ \emph {et~al.}(2010{\natexlab{b}})\citenamefont
  {Wu}, \citenamefont {Plapp},\ and\ \citenamefont {Voorhees}}]{Wu10b}%
  \BibitemOpen
  \bibfield  {author} {\bibinfo {author} {\bibfnamefont {K.-A.}\ \bibnamefont
  {Wu}}, \bibinfo {author} {\bibfnamefont {M.}~\bibnamefont {Plapp}}, \ and\
  \bibinfo {author} {\bibfnamefont {P.~W.}\ \bibnamefont {Voorhees}},\
  }\href@noop {} {\bibfield  {journal} {\bibinfo  {journal} {J. Phys.: Condens.
  Matter}\ }\textbf {\bibinfo {volume} {22}},\ \bibinfo {pages} {364102}
  (\bibinfo {year} {2010}{\natexlab{b}})}\BibitemShut {NoStop}%
\bibitem [{\citenamefont {Rechtsman}\ \emph {et~al.}(2005)\citenamefont
  {Rechtsman}, \citenamefont {Stillinger},\ and\ \citenamefont
  {Torquato}}]{Rechtsman05}%
  \BibitemOpen
  \bibfield  {author} {\bibinfo {author} {\bibfnamefont {M.~C.}\ \bibnamefont
  {Rechtsman}}, \bibinfo {author} {\bibfnamefont {F.~H.}\ \bibnamefont
  {Stillinger}}, \ and\ \bibinfo {author} {\bibfnamefont {S.}~\bibnamefont
  {Torquato}},\ }\href@noop {} {\bibfield  {journal} {\bibinfo  {journal}
  {Phys. Rev. Lett.}\ }\textbf {\bibinfo {volume} {95}},\ \bibinfo {pages}
  {228301} (\bibinfo {year} {2005})}\BibitemShut {NoStop}%
\bibitem [{\citenamefont {Engel}\ and\ \citenamefont {Trebin}(2007)}]{Engel07}%
  \BibitemOpen
  \bibfield  {author} {\bibinfo {author} {\bibfnamefont {M.}~\bibnamefont
  {Engel}}\ and\ \bibinfo {author} {\bibfnamefont {H.-R.}\ \bibnamefont
  {Trebin}},\ }\href@noop {} {\bibfield  {journal} {\bibinfo  {journal} {Phys.
  Rev. Lett.}\ }\textbf {\bibinfo {volume} {98}},\ \bibinfo {pages} {225505}
  (\bibinfo {year} {2007})}\BibitemShut {NoStop}%
\bibitem [{\citenamefont {Edlund}\ \emph {et~al.}(2011)\citenamefont {Edlund},
  \citenamefont {Lindgren},\ and\ \citenamefont {Jacobi}}]{Edlund11}%
  \BibitemOpen
  \bibfield  {author} {\bibinfo {author} {\bibfnamefont {E.}~\bibnamefont
  {Edlund}}, \bibinfo {author} {\bibfnamefont {O.}~\bibnamefont {Lindgren}}, \
  and\ \bibinfo {author} {\bibfnamefont {M.~N.}\ \bibnamefont {Jacobi}},\
  }\href@noop {} {\bibfield  {journal} {\bibinfo  {journal} {Phys. Rev. Lett.}\
  }\textbf {\bibinfo {volume} {107}},\ \bibinfo {pages} {085503} (\bibinfo
  {year} {2011})}\BibitemShut {NoStop}%
\bibitem [{\citenamefont {Edlund}\ \emph {et~al.}(2012)\citenamefont {Edlund},
  \citenamefont {Lindgren},\ and\ \citenamefont {Jacobi}}]{Edlund12}%
  \BibitemOpen
  \bibfield  {author} {\bibinfo {author} {\bibfnamefont {E.}~\bibnamefont
  {Edlund}}, \bibinfo {author} {\bibfnamefont {O.}~\bibnamefont {Lindgren}}, \
  and\ \bibinfo {author} {\bibfnamefont {M.~N.}\ \bibnamefont {Jacobi}},\
  }\href@noop {} {\bibfield  {journal} {\bibinfo  {journal} {Phys. Rev. Lett.}\
  }\textbf {\bibinfo {volume} {108}},\ \bibinfo {pages} {165502} (\bibinfo
  {year} {2012})}\BibitemShut {NoStop}%
\bibitem [{\citenamefont {Batten}\ \emph {et~al.}(2011)\citenamefont {Batten},
  \citenamefont {Huse}, \citenamefont {Stillinger},\ and\ \citenamefont
  {Torquato}}]{Batten11}%
  \BibitemOpen
  \bibfield  {author} {\bibinfo {author} {\bibfnamefont {R.~D.}\ \bibnamefont
  {Batten}}, \bibinfo {author} {\bibfnamefont {D.~A.}\ \bibnamefont {Huse}},
  \bibinfo {author} {\bibfnamefont {F.~H.}\ \bibnamefont {Stillinger}}, \ and\
  \bibinfo {author} {\bibfnamefont {S.}~\bibnamefont {Torquato}},\ }\href@noop
  {} {\bibfield  {journal} {\bibinfo  {journal} {Soft Matter}\ }\textbf
  {\bibinfo {volume} {7}},\ \bibinfo {pages} {6194} (\bibinfo {year}
  {2011})}\BibitemShut {NoStop}%
\bibitem [{\citenamefont {Lifshitz}\ and\ \citenamefont
  {Petrich}(1997)}]{Lifshitz97}%
  \BibitemOpen
  \bibfield  {author} {\bibinfo {author} {\bibfnamefont {R.}~\bibnamefont
  {Lifshitz}}\ and\ \bibinfo {author} {\bibfnamefont {D.~M.}\ \bibnamefont
  {Petrich}},\ }\href@noop {} {\bibfield  {journal} {\bibinfo  {journal} {Phys.
  Rev. Lett.}\ }\textbf {\bibinfo {volume} {79}},\ \bibinfo {pages} {1261}
  (\bibinfo {year} {1997})}\BibitemShut {NoStop}%
\bibitem [{\citenamefont {Cheng}\ and\ \citenamefont
  {Warren}(2008)}]{re:cheng08}%
  \BibitemOpen
  \bibfield  {author} {\bibinfo {author} {\bibfnamefont {M.}~\bibnamefont
  {Cheng}}\ and\ \bibinfo {author} {\bibfnamefont {J.~A.}\ \bibnamefont
  {Warren}},\ }\href@noop {} {\bibfield  {journal} {\bibinfo  {journal} {J.
  Comp. Phys.}\ }\textbf {\bibinfo {volume} {227}},\ \bibinfo {pages} {6241}
  (\bibinfo {year} {2008})}\BibitemShut {NoStop}%
\bibitem [{\citenamefont {Wu}\ and\ \citenamefont {Dzenis}(2008)}]{Wu08}%
  \BibitemOpen
  \bibfield  {author} {\bibinfo {author} {\bibfnamefont {X.-F.}\ \bibnamefont
  {Wu}}\ and\ \bibinfo {author} {\bibfnamefont {Y.~A.}\ \bibnamefont
  {Dzenis}},\ }\href@noop {} {\bibfield  {journal} {\bibinfo  {journal} {Phys.
  Rev. E}\ }\textbf {\bibinfo {volume} {77}},\ \bibinfo {pages} {031807}
  (\bibinfo {year} {2008})}\BibitemShut {NoStop}%
\bibitem [{\citenamefont {Mkhonta}\ \emph {et~al.}(2013)\citenamefont
  {Mkhonta}, \citenamefont {Vernon}, \citenamefont {Elder},\ and\ \citenamefont
  {Grant}}]{Mkhonta13}%
  \BibitemOpen
  \bibfield  {author} {\bibinfo {author} {\bibfnamefont {S.~K.}\ \bibnamefont
  {Mkhonta}}, \bibinfo {author} {\bibfnamefont {D.}~\bibnamefont {Vernon}},
  \bibinfo {author} {\bibfnamefont {K.~R.}\ \bibnamefont {Elder}}, \ and\
  \bibinfo {author} {\bibfnamefont {M.}~\bibnamefont {Grant}},\ }\href@noop {}
  {\bibfield  {journal} {\bibinfo  {journal} {EPL}\ }\textbf {\bibinfo {volume}
  {101}},\ \bibinfo {pages} {56004} (\bibinfo {year} {2013})}\BibitemShut
  {NoStop}%
\bibitem [{\citenamefont {Tegze}\ \emph {et~al.}(2011)\citenamefont {Tegze},
  \citenamefont {T\'oth},\ and\ \citenamefont {Gr\'an\'asy}}]{re:tegze11}%
  \BibitemOpen
  \bibfield  {author} {\bibinfo {author} {\bibfnamefont {G.}~\bibnamefont
  {Tegze}}, \bibinfo {author} {\bibfnamefont {G.~I.}\ \bibnamefont {T\'oth}}, \
  and\ \bibinfo {author} {\bibfnamefont {L.}~\bibnamefont {Gr\'an\'asy}},\
  }\href@noop {} {\bibfield  {journal} {\bibinfo  {journal} {Phys. Rev. Lett.}\
  }\textbf {\bibinfo {volume} {106}},\ \bibinfo {pages} {195502} (\bibinfo
  {year} {2011})}\BibitemShut {NoStop}%
\bibitem [{\citenamefont {Berry}\ \emph {et~al.}(2008)\citenamefont {Berry},
  \citenamefont {Elder},\ and\ \citenamefont {Grant}}]{re:berry08}%
  \BibitemOpen
  \bibfield  {author} {\bibinfo {author} {\bibfnamefont {J.}~\bibnamefont
  {Berry}}, \bibinfo {author} {\bibfnamefont {K.~R.}\ \bibnamefont {Elder}}, \
  and\ \bibinfo {author} {\bibfnamefont {M.}~\bibnamefont {Grant}},\
  }\href@noop {} {\bibfield  {journal} {\bibinfo  {journal} {Phys. Rev. E}\
  }\textbf {\bibinfo {volume} {77}},\ \bibinfo {pages} {061506} (\bibinfo
  {year} {2008})}\BibitemShut {NoStop}%
\bibitem [{\citenamefont {Chan}\ \emph {et~al.}(2010)\citenamefont {Chan},
  \citenamefont {Tsekenis}, \citenamefont {Dantzig}, \citenamefont {Dahmen},\
  and\ \citenamefont {Goldenfeld}}]{Chan10}%
  \BibitemOpen
  \bibfield  {author} {\bibinfo {author} {\bibfnamefont {P.~Y.}\ \bibnamefont
  {Chan}}, \bibinfo {author} {\bibfnamefont {G.}~\bibnamefont {Tsekenis}},
  \bibinfo {author} {\bibfnamefont {J.}~\bibnamefont {Dantzig}}, \bibinfo
  {author} {\bibfnamefont {K.~A.}\ \bibnamefont {Dahmen}}, \ and\ \bibinfo
  {author} {\bibfnamefont {N.}~\bibnamefont {Goldenfeld}},\ }\href@noop {}
  {\bibfield  {journal} {\bibinfo  {journal} {Phys. Rev. Lett.}\ }\textbf
  {\bibinfo {volume} {105}},\ \bibinfo {pages} {015502} (\bibinfo {year}
  {2010})}\BibitemShut {NoStop}%
\end{thebibliography}%

\end{document}